# An Energy Efficient Risk Notification Message Dissemination Protocol for Vehicular Ad hoc Networks

Natarajan Meghanathan
Assistant Professor
Department of Computer Science
Jackson State University
Jackson, MS 39217
Email: nmeghanathan@jsums.edu

Gordon W. Skelton
Associate Professor
Department of Computer Engineering
Jackson State University
Jackson, MS 39217
Email : gordon.skelton@jsums.edu

**Abstract**

The high-level contribution of this paper is the design and development of an energy-efficient Risk Notification Message Dissemination Protocol (RNMDP) for vehicular ad hoc networks (VANETs). RNMDP propagates Risk Notification Messages (RNMs) from their location of origin (called the Risk Zone) to vehicles approaching the Risk Zone. RNMDP assumes each node is aware of its current location in the network. The protocol works as follows: A RNM is broadcast in the neighborhood of the Risk Zone. A node receiving the RNM from another node waits for a Rebroadcast-Wait-Time before deciding to rebroadcast the message. The Rebroadcast-Wait-Time for a node is modeled based on the ratio of the distance between the node and the immediate sender of the RNM and the direction of movement of the node. Priority for rebroadcast is given for nodes farthest away from the sender and traveling towards the Risk Zone. Nodes that are traveling in lanes in direction away from the Risk Zone are also considered for rebroadcast, albeit with a larger Rebroadcast-Wait-Time. During the Rebroadcast-Wait-Time, if a node hears the same RNM again rebroadcast in the neighborhood, then the node stops from further broadcasting the message. If a node does not hear the RNM in its neighborhood during the Rebroadcast-Wait-Time, the node broadcasts the message in its neighborhood. A RNM is considered to have been delivered to all the vehicles in the road, if the message reaches the Target Zone. The performance of RNMDP has been compared with that of the commonly used flooding strategy through extensive simulations conducted for highway networks with different number of lanes and lane density. Simulation results indicate that with a slightly larger delay (i.e., no more than 35% of the delay incurred for flooding), RNMDP can achieve the same message delivery ratio attained by flooding, but at a relatively much lower energy loss compared to flooding.

**Key Words:** Broadcast, Vehicular Ad hoc Network (VANET), Message Dissemination, Energy, Latency

## 1. Introduction

Vehicular Ad hoc Networks (VANETs) are one of the most promising application areas of Mobile Ad hoc Networks (MANETs). A MANET is a dynamic distributed system of mobile wireless nodes. Communication in VANETs can be of two types: vehicle-to-vehicle communication and roadside-to-vehicle communication. VANET communication is normally accomplished through special electronic devices placed inside each vehicle so that an ad hoc network of the vehicles is formed on the road. A vehicle equipped with a VANET device should be able to receive and relay messages to other VANET-device equipped vehicles in its neighborhood.

One primary difference between a MANET and a VANET is in the mobility of the nodes involved. The mobility of the vehicles in VANETs is mostly constrained to follow a pattern (like the car following model [1], in which cars follow one after the other in a streamlined fashion), and is different from the random mobility of the nodes as seen in MANETs. Nevertheless, there are several common issues and concerns between MANETs and VANETs, and one of these primary concerns is energy consumption.

VANET devices are also battery powered (like MANET devices) and the source of power may be the gas in the fuel tank of the vehicle. It is of utmost importance that communication protocols developed for VANETs be energy sensitive and consume as much lower energy as possible.

One of the most common objectives of VANET communications is to contribute towards safer driving and provide timely information to drivers and other concerned authorities [2]. To accomplish this objective, it is imperative to develop broadcast message dissemination protocols for fast and reliable propagation of warning messages (in case of accidents, unexpected fog banks, dangerous road surface conditions and etc) to upcoming vehicles on the road. The alert messages should be delivered to the largest number of upstream vehicles in the shortest possible time. Flooding is one of the commonly used approaches in ad hoc networks to disseminate a message from one node to every other node in the network. But, flooding leads to significant amount of energy and bandwidth consumption as each node in the network is required to broadcast the message in its neighborhood [3].

In this paper, we propose a novel Risk Notification Message Dissemination Protocol (RNMDP) to propagate Risk Notification Messages (RNMs) from the location of origin (called the Risk Zone) to the vehicles approaching the Risk Zone. RNMDP has been designed to minimize the energy lost in the propagation of the RNMs, but at the same time incur the least possible delay in delivering the messages. RNMDP works as follows: The Risk Notification Message (RNM) originates in the Risk Zone and it must propagate towards vehicles approaching the Risk Zone. The location up to which the RNM has to propagate is called the Target Zone. Each vehicle on the road is assumed to be capable of determining its current location in the network (using location services like Global Positioning System, GPS [4]). All the vehicles are assumed to operate in a fixed transmission range. The RNM is broadcast by an initiating device in the Risk Zone in its neighborhood defined by the transmission range. Every node receiving the RNM waits for a time called the Rebroadcast-Wait-Time before deciding to rebroadcast the message. The value of the Rebroadcast-Wait-Time for a node depends on the distance of the node to the sender of the RNM (actually the ratio of the distance between these two nodes) and the direction of movement of the vehicle. During the Rebroadcast-Wait-Time, if a node hears the same RNM again in its neighborhood, the node does not rebroadcast the message. If a node does not hear the RNM in its neighborhood during the Rebroadcast-Wait-Time, the node broadcasts the message in its neighborhood defined by the transmission range. A RNM is considered to have been delivered to all the vehicles in the road, if the message reaches the Target Zone.

RNMDP incurs the least number of intermediate retransmitting nodes to rebroadcast the RNM. Hence, it incurs the least possible energy consumption. We use flooding as the benchmark to compare and evaluate the performance of RNMDP under different conditions of lane density and number of lanes. Performance simulation studies illustrate that the Rebroadcast-Wait-Time does not significantly increase the end-to-end delay for the messages delivered. RNMDP performs very well especially in high-density networks incurring the least possible energy consumption with a delay very closer to that incurred with flooding. Also, in networks of low lane density, RNMDP efficiently makes use of the increase in the number of lanes.

The rest of the paper is organized as follows: In Section 2, we discuss related work on broadcast protocols in VANETs. Section 3 proposes the RNMDP protocol and explains its working in detail. Section 4 presents the simulation environment and discusses the simulation results with respect to different simulation conditions. Section 5 concludes the paper and outlines the future work.

## 2. Related Work

The Smart Broadcast (SB) protocol proposed in [5] divides the rectangular forward area (called the area unit, AU) covered by the transmission range of a node into several sectors, controlled by a parameter identified as $N_S$. Each sector covers a fraction $1/N_S$ of an AU. The notification messages to be broadcast include the location information of the current transmitter of the message. Each node receiving the notification message chooses a rebroadcast-latency depending on the sector in which the node is located. Nodes lying in a sector far away from the sender will get the priority to rebroadcast the message. Some of the potential problems with the SB protocol are that all nodes lying with in the far-away sector will contend for channel access to rebroadcast the message and this can lead to unnecessary message collisions. The value of the parameter $N_S$ has to be dynamically chosen depending on the lane density and the number of lanes. Also, the SB protocol only allows nodes that are approaching the source of the broadcast to propagate the message downstream. This can lead to poor connectivity in low density networks. RNMDP does not have to deal with channel contention issues as only one node gets the highest priority to rebroadcast the message. RNMDP also allows nodes traveling in lanes in the direction away from the Risk Zone to get priority to rebroadcast a message, albeit with a higher Rebroadcast-Wait-Time. Thus, RNMDP can effectively deal with low lane density situations.

In the Distributed Vehicular Broadcast (DV-CAST) protocol [6], each node is assumed to periodically exchange beacons in its neighborhood to determine the status of the neighborhood, which could be categorized as: well-connected, sparsely-connected or totally disconnected. A node handles the broadcast message received depending on the status of its neighborhood. To avoid redundant broadcasts in a well-connected neighborhood, a broadcast suppression mechanism is used during which a node listens to the neighborhood for the number of redundant broadcasts received before deciding whether to propagate or not. Periodic beacon exchange is a core-requirement of DV-CAST and this can lead to high bandwidth and energy consumption. The RNMDP protocol proposed in this paper is beaconless, i.e., does not require the periodic exchange of beacons in the neighborhood.

In the Backfire algorithm proposed in [7], a node immediately rebroadcasts a message if it receives a message from a node that is separated from it by a distance greater than a threshold value of $D_{max}$. A node, whose distance to the originator is below the threshold $D_{max}$ value, waits for a time proportional to the distance to the originator. If a rebroadcast message is heard from some other node during this waiting time, the node decides to drop the message. The proper selection of $D_{max}$ value, depending on the node transmission range and lane density, is very critical for the Backfire algorithm to function effectively. RNMDP does not employ any such critical parameters that have to be differently configured in different operating scenarios.

## 3. RNMDP Protocol

In this section, we describe our novel energy-efficient Risk Notification Message Dissemination Protocol (RNMDP) to propagate a risk notification message from the point of origin, called the Risk Zone, in a road to vehicles approaching the Risk Zone.

### 3.1 Assumptions

RNMDP will use one dedicated frequency channel in the Intelligent Transport Systems, ITS, band of 5.85 – 5.925 GHz [8] and the channel will be well separated from the frequency channels used for comfort applications. A vehicle will have to tune to the RNMDP channel to receive messages corresponding to this protocol. In the rollout phase (the time period during which the majority of the vehicles are not designed to be compatible with the ITS band), where only one frequency channel may be used for all vehicular communications, RNMDP messages will be given more priority over messages belonging to other non-emergency, comfort applications. RNMDP assumes that all the vehicles using the protocol are GPS-enabled. Using Global Positioning System (GPS) [4], a vehicle can learn its own position on the road. For simplicity, in the description of the protocol in this paper, we assume a highway with two lanes in opposite directions. RNMDP will also work for highways with multiple lanes in same or opposite directions. The mobility model for the vehicles is the commonly used car-following model [1], according to which the speed of a vehicle is dependent on the speed of the vehicle ahead of it and the distance between the two vehicles.

### 3.2 Protocol Description

Let ($X_{min}$, $Y_{min}$, $X_{max}$, $Y_{max}$) be the boundaries of the area (hereafter called the "Risk Zone") about which the vehicles traveling towards it need to be aware of. The Risk Notification Message (RNM) is originated to spread any risk, safety threat or warning information for this area. The message could be initiated by a police vehicle on the road or a sink node gathering information from the sensors located near the Risk Zone. Each time a RNM is initiated near the Risk Zone, its contents could be different depending upon the up-to-date status of the Risk Zone. In Figure 1, node A is the initial sender of the RNM. Note that we use the words "node" and "vehicle" interchangeably. They are one and the same.

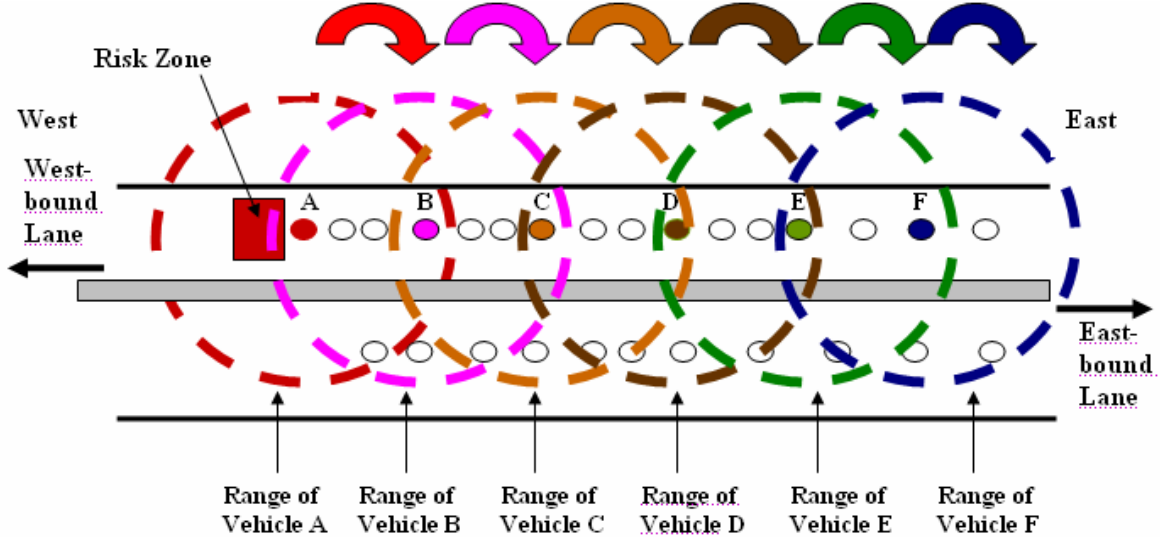

**Figure 1:** Propagation of the Risk Notification Message (RNM) away from the Risk Zone

#### 3.2.1 Structure of the Risk Notification Message

The structure of the Risk Notification Message is as shown in Figure 2. The message has 69 bytes of control overhead and the data portion of the message (the actual risk notification) is of size at most 512 bytes, typical value for medium-sized messages in ad hoc networks.

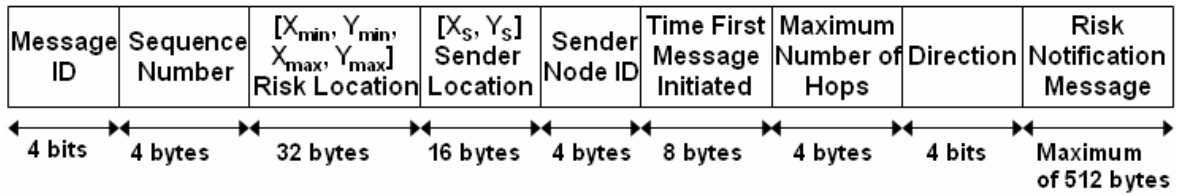

**Figure 2:** Structure of the Risk Notification Message (RNM) used by RNMDP

#### 3.2.2 Uniqueness of Risk Notification Messages

The Risk Notification Message is first broadcast around the neighborhood of the initiator node. A distinct value for the Message ID field would be used to indicate that the message is a Risk Notification Message and is different from messages belonging to other applications. The Risk Notification Messages are broadcast periodically until the Risk Zone is danger-free and the vehicles could pass through it without any harm. The time the first RNM is initiated due to the current risk at the Risk Zone is included in all the RNMs sent from that zone. The co-ordinates of the Risk Zone and the time of origin of the first RNM are together used to distinguish between RNMs originating at different locations at different time instants. RNMs from a particular Risk Zone with the same timestamp of origin are distinguished from one another using monotonically increasing sequence numbers.

### 3.2.3 Direction of Propagation of the Risk Notification Message

The direction the RNM should propagate is indicated using the 4-bit "Directions" field by using unique identifiers for each of the eight possible directions (North, North-east, East, South-east, South, South-west, West and North-west). If the message needs to be propagated in all directions, then all the 4 bits of the Directions field are set to 1. The risk information should be notified to the vehicles that approach the Risk Zone. So, the direction of propagation of the Risk Notification Message should be opposite to that of the direction of vehicles approaching the Risk Zone. Hence, the preference to rebroadcast would be given to the vehicles traveling towards the Risk Zone.

### 3.2.4 Maximum Number of Hops for Propagation

The initiator sets the maximum number of hops a Risk Notification Message could be broadcast in the "Maximum Number of Hops" field. The value of this field, as set by the initiator, is greater than zero. Each node receiving a RNM with a particular sequence number for the first time, decrements the *Maximum Number of Hops* field in the message by 1, before deciding to broadcast the message further. If the value of the *Maximum Number of Hops* field is equal to zero, then the node discards the RNM. Otherwise, i.e., if the value of the *Maximum Number of Hops* field is still greater than zero, the node waits for a "Rebroadcast-Wait-Time" before deciding whether to further rebroadcast the message or not.

### 3.2.5 Rebroadcast Wait Time

We do not want every node to forward the message because this would cause generation of redundant messages and message collisions, which may trigger the familiar broadcast storm problem [3]. Let $D$ denote the "Maximum Rebroadcast-Wait-Time", $R$ be the distance between a node and the node from which it received the Risk Notification Message and $R_{max}$ be the maximum transmission range of the nodes (which for simplicity, is the same for all nodes) in the network.

For nodes traveling towards the Risk Zone, the Rebroadcast-Wait-Time = $\frac{D}{2}\left[1 - \frac{R}{R_{max}}\right]$. During this Rebroadcast-Wait-time, if the node receives a RNM with a *Maximum Number of Hops* value that it has already seen, then the node does not rebroadcast the message by itself. If no such RNM is received within the Rebroadcast-Wait-time, the node decides to further rebroadcast the message in its neighborhood. The above equation illustrates that preference to rebroadcast is given to the neighbor node that is the farthest from the sender of the RNM. In Figure 1, node B would rebroadcast the RNM received from node A.

In order to make RNMDP robust and improve the chances of disseminating the RNM to the nodes approaching the Risk Zone, we also consider rebroadcasting through nodes traveling in lanes in the direction away from the Risk Zone. These nodes are made to wait for a Rebroadcast-Wait-Time = $\frac{D}{2} + \frac{D}{2}\left[1 - \frac{R}{R_{max}}\right]$ before propagation. Note that we still give first preference for rebroadcast by nodes traveling towards the Risk zone. Only if there are no such nodes to rebroadcast, especially in sparsely dense networks, nodes traveling away from the Risk zone are considered for rebroadcast. Again, among such nodes, a node that is traveling farthest away from the sender of the RNM is given preference.

### 3.2.6 Propagation of the Risk Notification Message

When a node decides to rebroadcast the Risk Notification Message after its Rebroadcast-Wait-Time expires, it updates the values of the Sender Location ($X_S$, $Y_S$) and Sender Node ID fields in the message using values that indicates itself. Note that the *Maximum Number of Hops* field would have been already decremented before the Rebroadcast-Wait-Timer was started. In Figure 1, node B would be the node that will rebroadcast the RNM in the neighborhood of node A. The above procedure is followed at every hop until the *Maximum Number of Hops* field value reaches zero. In Figure 1, the sequence of nodes that rebroadcast the RNM would be A, B, C, D, E and F, one node per hop count value.

### 4 Simulations

As a first step towards developing and evaluating RNMDP, we implemented the protocol in the ns-2 simulator [9]. We compare the performance of RNMDP with that of the conventionally used flooding algorithm to disseminate messages from one node to every other node in an ad hoc network. We conducted simulations with networks depicting highways of dimensions 5m x 8000m (refer Figure 3) and 10m x 8000m (refer Figure 4). The highway of dimensions 5m x 8000m is unidirectional and has only one lane, while the highway of dimensions 10m x 8000m is bi-directional and has two lanes, in opposite directions. The width of each lane is 5m for both the highways. Note that 'm' as a unit notation corresponds to meters and not miles. For each highway, the Risk Zone is assumed to be located at (0, 0) and the location up to which the risk notification message has to propagate (hereafter referred to as the Target Zone) is assumed to be at (5, 5000). We follow the 3-seond rule for minimum inter-vehicular spacing in the highways. Accordingly, the minimum distance between two vehicles (referred as the safe distance) in a lane is 92 meters (300 ft), coinciding with an Interstate vehicular speed of 65-70 miles per hour.

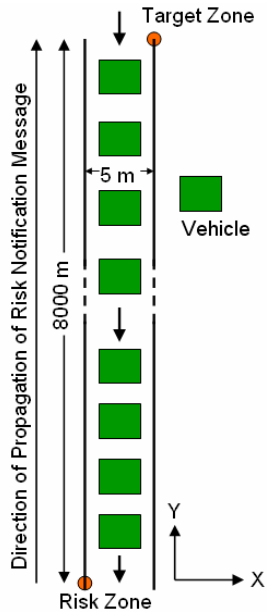 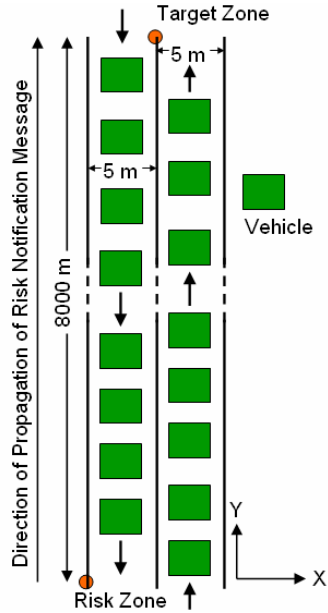

**Figure 3:** One-lane Highway        **Figure 4:** Two-lane Highway

Simulations are conducted for highways with three different lane density values. Highways with 30 nodes per lane, 45 nodes per lane and 60 nodes per lane are referred to as low, medium and high lane density networks. Each vehicle moves with a speed uniformly distributed between 65 to 70 miles per hour. The mobility model used is the widely used car-following model [1]. The MAC layer model followed is the IEEE 802.11 model [10]. The signal propagation model is the Two-ray Ground Reflection Model [10]. Each node is location-aware and can identify its location in the network at any point of time. The energy consumed [11] due to broadcast transmission of a message within the neighborhood is 1.1182 + $7.2*10^{-11}*R^4$ where $R$ is the transmission range of a node. The energy consumed [11] to receive a message is 1W. We only consider the energy losses due to transmission and reception.

For both the highways, the number of risk notification messages sent from the Risk Zone to the Target Zone is 4000. The messages are sent at the rate of 1 message per second, periodically one after another. The message size is 512 bytes. The channel bandwidth is assumed to be 2 Mbps. Simulations are conducted for different transmission range per node values: 250m to 1000m, in increments of 50m. For a given simulation, the transmission range of all the nodes is fixed and is the same. Each node has a queue of size 200 (i.e., the queue can hold 200 messages at any point of time) and the queue works in First-In-Out-First fashion. We assume that the propagation of the RNM is the only event happening in the network considered and that there is no other data exchange event taking place.

The value of the Maximum-Rebroadcast-Wait-Time (parameter $D$ in the Rebroadcast-Wait-Time model described in Section 3.2.5) is 1 sec. With this $D$ value, the maximum value for the Rebroadcast-Wait-Time for nodes traveling towards the Risk Zone and the minimum value for the Rebroadcast-Wait-Time for nodes traveling away from the Risk zone are both 0.5 seconds each. Typical Rebroadcast-Wait-Time values for nodes traveling towards the Risk Zone are in the 0.2-0.35 seconds range and for nodes traveling away from the Risk Zone are in the 0.7-0.85 seconds range.

We assume that there exists a transmitting device (like a laptop in the police vehicle or a sink node gathering information from the sensors) in the vicinity of the Risk Zone. This transmitting device is the initiator of the Risk Notification Message. This message has to be broadcast to all the vehicles approaching the Risk Zone. We also assume that there exists a node in the Target Zone. The node in the Target Zone has to receive the RNM in order for us to conclude that the message propagated the entire region of the highway lane starting from the Risk Zone to the Target Zone and that every other node in this region received the message.

**4.1 Flooding**

In this section, we give an overview of the flooding algorithm which has been used as the benchmark to evaluate the performance of RNMDP in the simulations. Flooding is a widely-used approach for disseminating a message from one node to all the nodes in a network. In the case of on-demand ad hoc routing protocols [12][13], flooding has been also used to discover a path between a pair of nodes in the network, whenever required. For a given network density, flooding offers the highest probability for each node in the network to receive one or more copies of the flooded message.

We simulate flooding as follows: The initiating node sets up monotonically increasing sequence number, origin time, location of the Risk Zone and other necessary information in the header (refer Figure 5) of the notification message. The initiator sets the *Maximum Number of Hops* field to control the maximum number of hops the notification message can travel. The message is then broadcast in the neighborhood of the initiator. All nodes that are within the transmission range of the initiator are said to be in the neighborhood of that node. Each node that receives the Risk Notification Message decrements the *Maximum Number of Hops* field in the message by 1 and decides to broadcast the message in its neighborhood (i) if the value of the *Maximum Number of Hops* field in the message after the decrement operation is still greater than zero and (ii) if the node has not seen a message with *Sequence Number* higher or equal to that received. Thus, with flooding, each node in the network broadcasts a Risk Notification Message with a particular sequence number exactly once. The receiving node in the Target Zone may get one more copy of the Risk Notification Message as it propagates through several paths in the network. All the fields in the header of the Risk Notification Message shown in Figure 5 are required in order for the message to be informative to the receivers and for flooding to function effectively and efficiently as explained above. The size of the message header used for flooding is 48 bytes and 4 bits long, which is only 20 bytes and 4 bits less than the size of the message header used by RNMDP.

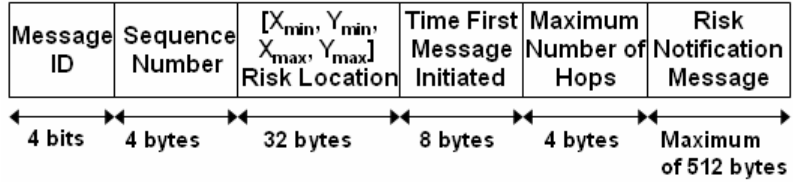

**Figure 5:** Structure of the Risk Notification Message used for Flooding

### 4.2 Performance Metrics

We measure the following performance metrics through the simulations:
- Message Delivery Ratio: This metric is the ratio of the number of messages received at the Target Zone to the number of messages transmitted from the Risk Zone. The Message Delivery Ratio indicates the probability of the Target Zone connected to the Risk Zone through one or more intermediate nodes.
- Number of Retransmitting Nodes: This metric refers to the number of intermediate nodes broadcasting (i.e., retransmitting) the Risk Notification Message as the message propagates from the Risk Zone to the Target Zone.
- Energy Lost per Message Delivered: This metric is the sum of the transmission energy and receiving energy consumed at every node in the network during the broadcast of the Risk Notification Message from the Risk Zone to the Target Zone
- Delay per Message Delivered: This metric is the end-to-end delay suffered by a message when it reaches the Target Zone. The delay suffered by the message includes the propagation delay and transmission delay at each hop and the channel access delay and the queuing delay spent at the queues of the nodes. In the case of RNMDP, the delay suffered by a message also includes the Rebroadcast-Wait-Time spent by the message at each intermediate forwarding node.

For a given simulation condition, the performance metrics are collected by running the simulation with 5 samples of the mobility profile of the nodes and then averaging the results.

### 4.3 Message Delivery Ratio

The message delivery ratio is a measure of network connectivity. For a given simulation condition, we obtain identical message delivery ratios for both flooding and RNMDP. For the highway with only one lane, the message delivery ratio reaches unity only when the lane density is moderate and high (refer Figure 6). The minimum transmission range per node values at which this is obtained are 400m and 800m respectively for high (60 nodes per lane) and moderate (45 nodes per lane) lane densities. For networks of low lane density (i.e., with only 30 nodes in the lane), the message delivery ratio is more than 0.9 only when the transmission range per node is 1000m.

For the highway with two lanes, similar to the highway with one lane, the message delivery ratio reaches unity only when the lane density is moderate and high (refer Figure 7). The minimum transmission range per node values at which this is obtained are 350m and 750m respectively for high (60 nodes per lane) and moderate (45 nodes per lane) lane densities. Thus, for high and moderate lane density networks, with the addition of another lane, there is no significant reduction in the minimum transmission range per node value to ensure 100% message delivery. Nevertheless, for low lane density networks (30 nodes per lane), we observe that for highways with two lanes, the message delivery ratio is more than 0.9 for transmission range per node values starting from 750m. Thus, adding a second lane is relatively more effective for networks with low lane density.

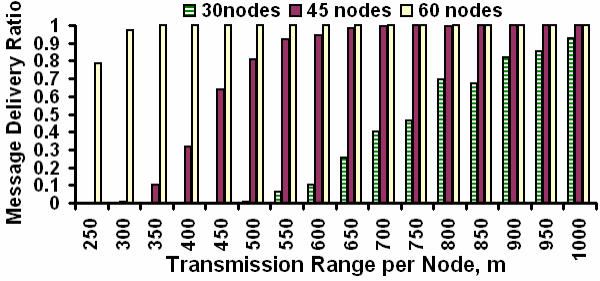
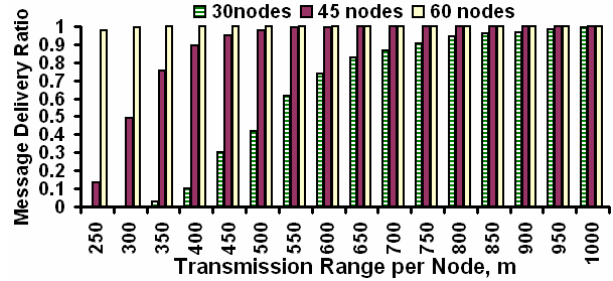

**Figure 6:** Message Delivery Ratio (Highway with One Lane)

**Figure 7:** Message Delivery Ratio (Highway with Two Lanes)

### 4.4 Number of Retransmitting Nodes

In the case of RNMDP, for a given lane density, as we increase the number of lanes from one lane to two lanes, the number of retransmitting nodes relatively decreases, especially when the transmission range per node values are low. This is attributed to the increase in the connectivity of the network and only nodes that are relatively far away from each other need to rebroadcast. The number of retransmitting nodes in highway with only one lane is at most 15% (high lane density), 12% (moderate lane density) and 10% (low lane density) more than the number of retransmitting nodes in highway with two lanes. On the other hand, in the case of flooding, for a given lane density, as the number of lanes is doubled from one to two, the number of retransmitting nodes also doubles. This is due to the requirement of flooding that every intermediate node in the network should broadcast a message exactly once. With RNMDP, the message gets propagated through only one path from the Risk Zone to the Target Zone; where as with flooding, the message gets propagated through several paths from the Risk Zone to the Target Zone. RNMDP effectively makes use of the highly one-dimensional nature of the highway networks and achieves the objective of delivering the Risk Notification Message to every node in the region between the Risk Zone and Target Zone by just propagating the message through only one path in this region.

For highways with moderate and high lane density, as we increase the transmission range per node value from 250m to 1000m, (i.e., as the transmission range per node value is quadrupled), the number of retransmitting nodes is reduced to one-fourth or one-fifth. This inverse linear relationship between the transmission range per node and the number of retransmitting nodes can be attributed to the highly one-dimensional nature of highway networks. In such networks, for any two nodes, if the magnitude of the difference in their co-ordinate values in the larger dimension is less than or equal to the transmission range per node, then the two nodes are likely to be located within the transmission range of each other. The magnitude of the difference in the co-ordinate values in the smaller dimension is almost negligible.

### 4.5 Energy Lost per Message Delivered

This metric refers to the average energy lost in the network as a Risk Notification Message propagates in the network from the Risk Zone to the Target Zone. The energy lost at a node is mainly due to the broadcast transmission of the message in the neighborhood and reception of the notification message from the nodes in the neighborhood. First, we discuss the energy lost per message delivered for RNMDP considering the highways with one lane and two lanes and different lane density scenarios (refer Figures 8 and 9). We will then compare the energy lost per message delivered for RNMDP and flooding (refer Figures 10 and 11).

#### 4.5.1   RNMDP: Energy Lost per Message Delivered

In the case of RNMDP, to achieve 100% or the maximum possible message delivery, the number of nodes required to retransmit decreases as the transmission range per node value increases. More nodes may be required to retransmit if the transmission range per node value is lower. But, as observed in the results for energy consumption (refer Figures 8 and 9), the larger the transmission range, the more the energy consumed to guarantee 100% or the maximum possible message delivery. Hence, for a given highway and lane density, it would be wise to operate at a transmission range (referred to as the critical transmission range) such that the network incurs the lowest energy consumption, but still achieves 100% or the maximum possible message delivery ratio. For both high and moderate lane density networks, as we increase the number of lanes from one to two, the critical transmission range decreases by 12%. However, at these critical transmission range values, the energy lost per message delivered in highway with two lanes is 5% and 70% more than the energy lost per message delivered in highway with one lane at moderate and high lane densities respectively. For the low lane density network considered for the simulations, we could not achieve 100% message delivery and could achieve only a maximum delivery ratio less than 1 even at the maximum transmission range value of 1000m. The energy lost per message delivered in highway with two lanes in low lane density networks is only 8-10% more than the energy lost per message delivered in highway with one lane.

For a given lane density, as we increase the number of lanes from one to two, the increase in the energy lost per message delivered is by a factor of 70% (for low lane density highway) to 90% (for high lane density highway) at transmission range per node values equal to or less than 400m and is by a factor of 7% (for low lane density highway) to 20% (for high lane density highway) at transmission range per node values greater than 900m. As the transmission range per node increases, the decrease in the rate of increase of energy lost per message delivered is attributed to the reduction in the number of hops the message travels from the Risk Zone to the Target Zone. As we increase the transmission range per node value from 250m to 1000m, the average hop count per path traversed by RNM decreases by about 75% to 80%. When we consider only the network scenarios in which there is a non-zero message delivery ratio, for a given lane density and transmission range per node value, there is no appreciable difference in the average hop count of the path traversed by the RNM.

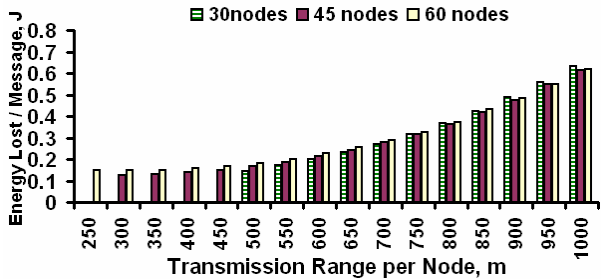 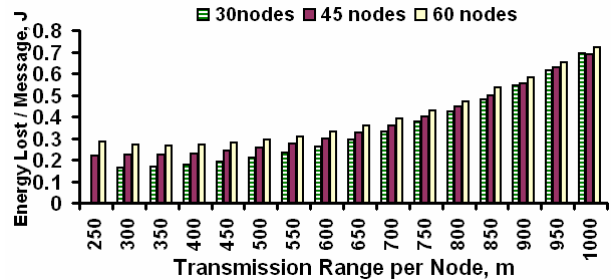

**Figure 8:** RNMDP – Energy Lost per Message (Highway with One Lane)  **Figure 9:** RNMDP – Energy Lost per Message (Highway with Two Lanes)

For highway with two-lanes, as we increase the lane density from low to moderate and to high, the increase in the energy lost per message delivered is by a factor of 50-70% (as we increase from low to high lane density) and 17-22% (as we increase from moderate to high lane density) at transmission range per node values less than or equal to 400m and is by a factor of 4 to 7% (as we increase from low to high and from moderate to high lane densities) at transmission range per node values greater than or equal to 900m. A similar observation can also be made for highway with one lane, albeit the increase in the energy lost per message delivered is relatively lower as we increase the lane density from low to moderate and to high. The decrease in the relative increase in the energy lost per message delivered is attributed to the following two factors: (i) As we increase the transmission range per node value, the average hop count per path decreases and (ii) For a given transmission range per node value and number of lanes, as we increase the lane density from low to high, the average hop count per path decreases by 10-15% due to the increase in the number of neighbors per node and the node that is farthest away from the sender in the neighborhood manages to rebroadcast the message.

### 4.5.2 Flooding Vs RMNDP: Energy Lost per Message Delivered

RMNDP incurs the lowest energy lost per message delivered for all the scenarios simulated (refer Figures 10 and 11). This is justified by the relative reduction in the number of nodes retransmitting the Risk Notification Message compared to flooding. For a given lane density, as the transmission range per node value is increased from 250m to 1000m, energy lost due to flooding increases significantly. Similarly, for a fixed transmission range per node, with increase in the lane density and/or the number of lanes, energy lost due to flooding increases significantly. This can be attributed to the requirement of flooding that each node has to transmit the RNM exactly once, irrespective of the transmission range.

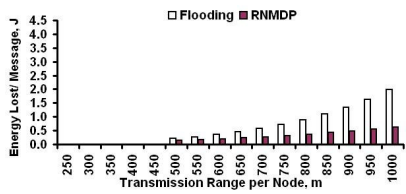 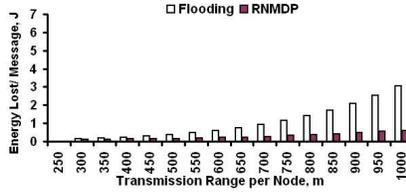 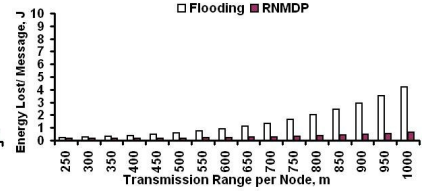

**Figure 10.1:** 30 Nodes   **Figure 10.2:** 45 Nodes   **Figure 10.3:** 60 Nodes

**Figure 10:** RNMDP vs. Flooding – Energy Lost per Message Delivered (Highway with One Lane)

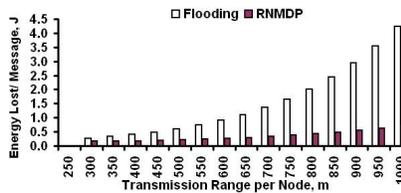 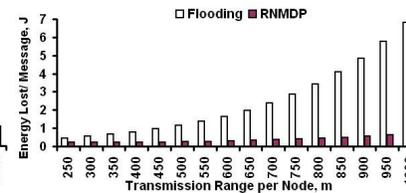 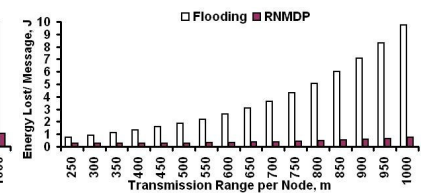

**Figure 11.1:** 30 Nodes   **Figure 11.2:** 45 Nodes   **Figure 11.3:** 60 Nodes

**Figure 11:** RNMDP vs. Flooding – Energy Lost per Message Delivered (Highway with Two Lanes)

For a given lane density, in both one lane and two lane scenarios, the difference in the magnitude of the energy lost due to RNMDP and flooding increases as we increase the transmission range per node. This is attributed to the fact that in the case of RNMDP, as the transmission range per node increases, the number of hops traversed by the RNM decreases significantly. Also, for a fixed number of lanes and transmission range per node, as the lane density increases, the hop count per path for RNMDP decreases slightly. For highway with two lanes, the energy lost per message delivered for RNMDP is about 60% (low lane density), 40-50% (moderate lane density) and 30-40% (high lane density) of the energy lost due to flooding at transmission range values less than or equal to 300m and is about 18-20% (for low lane density) and 8-12% (moderate and high lane density) of the energy lost due to flooding at transmission range values greater than or equal to 900m. For highway with one lane, the energy lost per message delivered for RNMDP is about 65-80% (for moderate lane density) and 57-75% (high lane density) of the energy lost due to flooding at transmission range values less than or equal to 300m and is about 36% (for low lane density), 22% (for moderate lane density) and 16% (for high lane density) of the energy lost due to flooding at transmission range values greater than or equal to 900m.

### 4.6 Delay per Message Delivered

This metric refers to the average of the end-to-end delays incurred by a Risk Notification Message from its point of origin at the Risk Zone to its final destination, which is the Target Zone. The end-to-end delay per message is the sum of the propagation delay and transmission delay at each hop and the channel access delay and queuing delay at each intermediate node that forwards the message from the Risk Zone to the Target Zone. We first discuss the delay per message delivered for RNMDP by considering the highways with one lane and two lanes and different lane density scenarios (refer Figures 12 and 13). We then compare the delay per message delivered for RNMDP vis-à-vis flooding (refer Figures 14 and 15).

### 4.6.1 RNMDP: Delay per Message Delivered

In low lane density scenarios, for a given transmission range per node value, as we increase the number of lanes from one to two, the increase in the end-to-end delay is by a factor of 5-8% for transmission range per node values less than or equal to 750m and is only by 2-3% for transmission range per node values above 750m. At the same time, we can observe that the message delivery ratio increases by 40-60% when we add a second lane. The slightly larger delay for highway with two lanes compared to the highway with one lane can be attributed to the slightly larger Broadcast-Wait-Time suffered by the messages when they get transmitted by nodes that are far away from each other. For moderate lane density scenarios, even though the message delivery ratio increases significantly with the introduction of the second lane, especially for transmission range per node values below 500m, there is no significant difference in the delay suffered by the messages. At high lane density scenarios, the message delivery ratio improved only marginally, at most by 20% for transmission range per node values below 300m, and the delay per message delivered almost remained the same for both highway with one lane and highway with two lanes scenarios.

For a given highway (one lane or two lanes) and a fixed transmission range per node value, the end-to-end delay per message decreases with increase in lane density. This can be attributed to the relatively lower hop count incurred by the RNM in networks of higher and moderate lane density. For highway with one lane and a fixed transmission range per node value, the end-to-end delay per message in high lane density networks is 75 to 83% of the delay incurred in low lane density networks and is about 88-95% of the delay incurred in moderate density networks. For highway with two lanes and with a fixed transmission range per node value, the end-to-end delay per message in high lane density networks is about 70-80% of the delay incurred in low lane density networks and is about 85-93% of the delay incurred in moderate lane density networks.

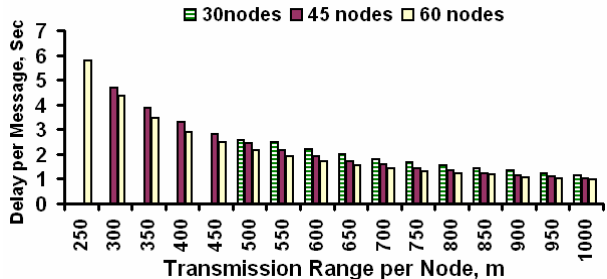
**Figure12:** RNMDP – Delay per Message
(Highway with One Lane)

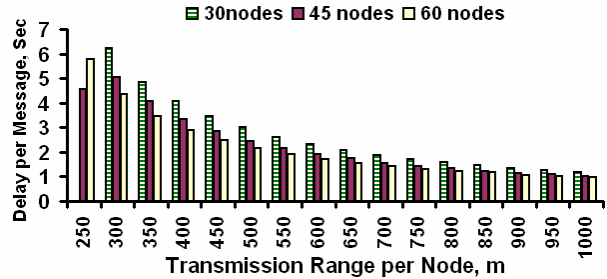
**Figure 13:** RNMDP - Delay per Message
(Highway with Two Lanes)

For a given lane density and highway (one lane or two lanes), as we increase the transmission range per node value, we observe a decrease in the end-to-end delay per message. But the decrease is at a faster rate as the transmission range value is increased up to a moderate value of 500-600m. With further increase in the transmission range per node value, the rate of decrease in the delay per packet is slower. This can be attributed to larger channel access delays and the larger queuing delays at the nodes in networks of larger neighborhood size. For example, in a single lane and high lane density highway, the end-to-end delay per data packet incurred at transmission range per node values of 250m, 500m and 1000m are 5.79 seconds, 2.16 seconds and 0.96 seconds respectively.

### 4.6.2 Flooding Vs RNMDP: Delay per Message Delivered

In the case of flooding, the Risk Notification Messages propagate through multiple different paths from the Risk Zone to the Target Zone. Hence, these RNMs are bound to incur different delays depending upon the path taken. However, for the purpose of measuring the end-to-end delay per message delivered, we take into consideration only the end-to-end delay incurred by the first Risk Notification Message reaching the Target Zone. All our discussions related to delay incurred with flooding for the rest of this paper are based on the end-to-end delay measured as stated above. The first RNM to reach the Target Zone follows the minimum hop path to the Target Zone and hence the end-to-end delay suffered by the RNMs in the case of flooding is almost always proportional to the hop count incurred.

For a given lane density and transmission range per node, we observe that the end-to-end delay suffered by the first message reaching the Target Zone using the flooding approach is independent of the number of lanes in the highway. For a given highway (one lane or two lanes) and for a given transmission range per node, as we increase the lane density from low to moderate and from low to high, the end-to-end delay per message delivered using flooding decreases by 7-8% and 10-14% respectively. Similar to RNMDP, we observe in the case of flooding, for a given lane density and highway (one lane or two lanes), as we increase the transmission range per node value, the end-to-end delay per message decreases, albeit at a faster rate up to moderate transmission range values of 500-600m and at a slower rate for larger transmission range values.

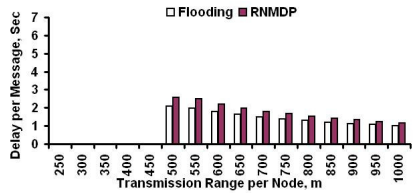
**Figure 14.1:** 30 Nodes
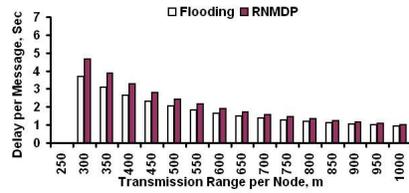
**Figure 14.2:** 45 Nodes
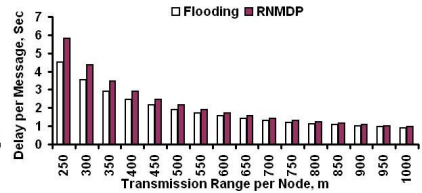
**Figure 14.3:** 60 Nodes

**Figure 14:** RNMDP vs. Flooding – Delay per Message Delivered (Highway with One Lane)

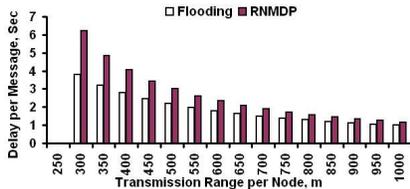
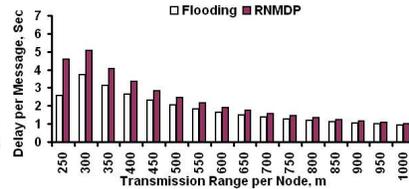
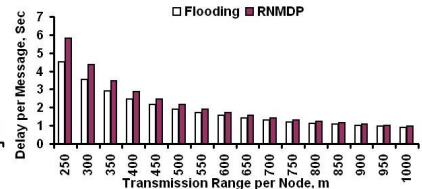
**Figure 15.1:** 30 Nodes
**Figure 15.2:** 45 Nodes
**Figure 15.3:** 60 Nodes

**Figure 15:** RNMDP vs. Flooding – Delay per Message Delivered (Highway with Two Lanes)

RNMDP incurs a relatively larger delay compared to flooding for all the simulated scenarios (refer Figures 14 and 15). This is understandable because the Risk Notification Message suffers a Rebroadcast-Wait-Time amount of delay at each node before being attempted for transmission from the node. The value of the Rebroadcast-Wait-Time is highly variable (as explained in Section 3.2.5) and is not fixed per node. Nevertheless, we observe that RNMDP suffers an end-to-end delay per message that is no more than 35% of the delay suffered by flooding in almost all the scenarios. The only scenarios for which the RNMDP end-to-end delay per message delivered is greater than its flooding counterpart are: highway with two lanes (45 nodes – transmission range per node of 250m, 30 lanes – transmission range per node of 300m to 450m). This is attributed to the relatively larger Rebroadcast-Wait-Time incurred by RNMDP for networks of very low lane density. Nevertheless, even in these scenarios, RNMDP effectively makes use of the addition of the second lane to improve the message delivery ratio compared to highway with single lane.

## 5   Conclusions and Future Work

RNMDP incurs relatively much lower energy consumption for delivery of the Risk Notification Messages compared to that incurred with flooding. The protocol does not require periodic exchange of beacons in the neighborhood and hence it also conserves network bandwidth. Performance simulation studies illustrate that the Rebroadcast-Wait-Time does not significantly increase the end-to-end delay for the messages delivered. So, even though, one can visualize a delay-energy consumption tradeoff between RNMDP and flooding, the tradeoff is not equal in both directions. With a slightly larger delay (i.e., no more than 35% of the delay incurred for flooding), RNMDP can achieve the same message delivery ratio attained by flooding, but at a relatively much lower energy loss compared to flooding. RNMDP performs very well especially in high-density networks incurring the least possible energy consumption with a delay very closer to that incurred with flooding. Also, in networks of low lane density, RNMDP efficiently makes use of the increase in the number of lanes by allowing nodes traveling in lanes in the direction away from the Risk Zone to get a chance to rebroadcast the RNMs, albeit with a higher Rebroadcast-Wait-Time. RNMDP does not have to deal much with channel contention issues as only one node gets the highest priority to rebroadcast the message at any point of time. Also, unlike other broadcast optimization protocols for VANETs, RNMDP does not employ any critical operating parameters that have to be differently configured in different network scenarios. Future work involves comparing RNMDP with VANET broadcasting protocols other than flooding. We are also planning for a real-time implementation of RNMDP in the near future.